# Structure of Velocity Distribution of Sheath-Accelerated Secondary Electrons in Asymmetric RF-DC Discharge

Alexander V. Khrabrov[1], Igor D. Kaganovich[1], Peter L. G. Ventzek[2], Alok Ranjan[3], and Lee Chen[2]

[1]*Princeton Plasma Physics Laboratory, Princeton NJ 08543*

[2]*Tokyo Electron America, Austin TX 78741*

[3]*Tokyo Electron Technology Center America, Albany NY*

(Wednesday, February 25, 2015)

Low-pressure capacitively-coupled discharges with additional DC bias applied to a separate electrode are utilized in plasma-assisted etching for semiconductor device manufacturing. Measurements of the electron velocity distribution function (EVDF) of the flux impinging on the wafer, as well as in the plasma bulk, show a thermal population and additional peaks within a broad range of energies. That range extends from the thermal level up to the value for the "ballistic" peak corresponding to the bias potential. The non-thermal electron flux has been correlated to alleviating the electron shading effect and providing etch-resistance properties to masking photoresist layers. "Middle-energy peak electrons" at energies of several hundred eV may provide an additional sustaining mechanism for the discharge. These features in the electron velocity (or energy) distribution functions are possibly caused by secondary electrons emitted from the electrodes and interacting with two high-voltage sheaths: a stationary sheath at the DC electrode and an oscillating, self-biased sheath at the powered electrode. Since at those energies the mean free path for large-angle scattering (momentum relaxation length) is comparable to, or exceeds the size of the discharge gap, these "ballistic" electrons will not be fully scattered by the background gas as they traverse the inter-electrode space. We have performed test-particle simulations where the features in the EVDF of electrons impacting the RF electrode are fully resolved at all energies. An analytical model has been developed to predict existence of peaked and step-like structures in the EVDF. These features can be explained by analyzing the kinematics of electron trajectories in the discharge gap. Step-like structures in the EVDF near the powered electrode appear due to accumulation of trapped electrons during a part of the RF cycle and their subsequent release. Trapping occurs when the RF sheath voltage exceeds the applied bias, and is decreasing. The step structures are formed by secondary electrons originating from the DC-biased surface (which also produce a peak near the energy equal to the bias potential). Additional peaks, at lower energies, are formed by the electrons emitted from the RF electrode and eventually escaping to it. The latter electrons can be grouped according to the number of bounces between the sheaths during their residence time in the discharge. Each of such groups may give rise to an individual peak in the distribution. The trap-and-release theory developed in this paper provides a convincing explanation for the observations of the ballistic and "middle energy peak" electrons measured in experiments.

## I. Introduction

Radio-frequency (RF) capacitively-coupled discharges find many important uses in plasma-assisted technologies. Among these technologies are plasma etching and deposition for fabrication of semiconductor devices in which patterned silicon wafers are exposed to reactive halogen-containing plasmas. Various configurations of such discharges exist, including those that augment the conventional RF source. In particular, it has been known [1] [2] that injecting energetic electron beam allows greater control and optimization of the plasma properties for such applications as etching and deposition. Early studies focused on optimizing plasma uniformity across the substrate being processed. Kushner *et al.* [3] found that the uniformity was influenced by direct injection of an electron beam into the plasma. In their study the beam was parallel to the wafer and did not interact directly with it. Shaw *et al.* [1] characterized the electron beam flux incident upon an unbiased electrode opposite to a DC/RF electrode in an inductively coupled plasma source. The electron beam in that experiment was produced by secondary emission from the DC-biased surface. Direct interactions of electron beam with surfaces have also been addressed in several works. Li and Joy [4] demonstrated the enhancement of film growth rates due to processes initiated by electron beam. Chung *et al.* [5] showed how electron impact affects roughness of the line width in experiments where 193 nm photoresist films were exposed concurrently to ions, electrons, and UV flux. The RF-DC configuration where both the powered and DC-biased electrodes are co-planar with the substrate, with secondary electrons emitted from the electrode opposite the wafer, has also been studied extensively, both in experiments and by numerical means [6] [7] [8] [9] [10]. The simple physical picture is that the secondary electrons accelerated through the DC sheath form a beam (in the velocity space) which then interacts with the plasma and with the substrate positioned opposite to the biased electrode.

While practical processes of plasma etching and deposition involve reactive gas mixtures (typically carbon- or halogen-containing), much of the essential physics can be tested with inert gases, such as argon, as a model system. Such was the case in the experiments by Chen and Funk [10] and by Chen *et al.* [11]. Previous studies focused on the energy distribution of the beam-like electron population produced in the DC sheath and impinging upon the substrate, and its effect on the plasma-material interaction. Important features of the electron distribution function were measured, inferred, or reproduced numerically. At the same time, a theoretical description of how the structure of the EVDF depends upon basic parameters of the DC/RF plasma source has yet to be presented in the literature.

The observation common among all studies is that the electron population within the plasma volume, and the one reaching the wafer, can be divided into a thermal group, a distinct intermediate-energy group, and a ballistic (largely unaffected by collisions) group which retains the energy and velocity-direction anisotropy imparted by the high-voltage sheaths. Each group has an identifiable role. The thermal (or "bulk") group contributes to the charge density and sets the plasma potential. The intermediate energy range is of importance because this is where electron impact cross-sections are near their maximum. The high-energy ballistic group, which is the focus of the present work, may penetrate into surface features or into subsurface layers of materials.

In this paper, we develop an idea that experimentally observed peaks in the velocity distribution of the secondary electrons could be due to the well-known effect of phase bunching, also long known as the klystron effect [12]. An alternative mechanism proposed in Ref. [10] involves a wave-particle interaction. It is well known that wave-particle interactions mediated by various kinds of instabilities can modulate the energy of an electron beam. Such collective mechanisms come into play for large beam currents, mono-energetic beams, and low pressures. They will be addressed in companion publications.

Secondary electrons emitted from the electrode surfaces accelerate through the sheaths and, depending on the profile of the electric potential, bounce between them a number of times, to form a temporarily trapped energetic population. These electrons can be an important source of gas ionization while trapped, and wafer charging when "dumped" onto the wafer. Trapping and dumping would certainly occur in any test-particle or self-consistent kinetic model of the discharge, but nature and origin of the corresponding EVDF structure is obscured in most analyses. Quantitative results can be obtained by examining the electron motion in the discharge gap.

We present an analysis of the kinematics of electron trajectories to understand the formation of the EVDF of the energetic group. The analysis is supported by simulations in which the cycle-averaged velocity distribution is fully resolved. To achieve the necessary resolution, the statistical fluctuations were reduced to a sufficiently small level by increasing the number of electrons to show the detailed structure of the EVDF at all energies. We show that kinematic analysis, the primary focus of which is the bunching effect, is sufficient to predict complex structures (peaks and steps) that could be experimentally observed in the distributions. The analysis does not exclude the possibility of wave-particle interactions playing a significant role in the EVDF formation.

This work is organized as follows. We use the experimental studies by Xu *et al*. [6] and by Chen and Funk [10] as the basis for comparison with the numerical results. The paper begins with a description of the experimental configuration, and then the numerical model. Examples of velocity distributions from test-particle Monte Carlo model are presented. Following this are theoretical interpretations of the numerical experiments for both collisionless and collisional cases.

## II. RF-DC Discharge Configuration

A schematic diagram of the experimental RF-DC discharge device of Ref. [6] is reproduced in Fig. 1. The RF current is blocked in the DC bias circuit and returns through the grounded side surface of the device. In the experiment, the area of that surface is ~5 times larger than that of the powered electrode, resulting in a highly asymmetric, self-biased RF configuration. An example of the electron distributions measured in the experiment from Ref. [6] are included in Fig. 1.

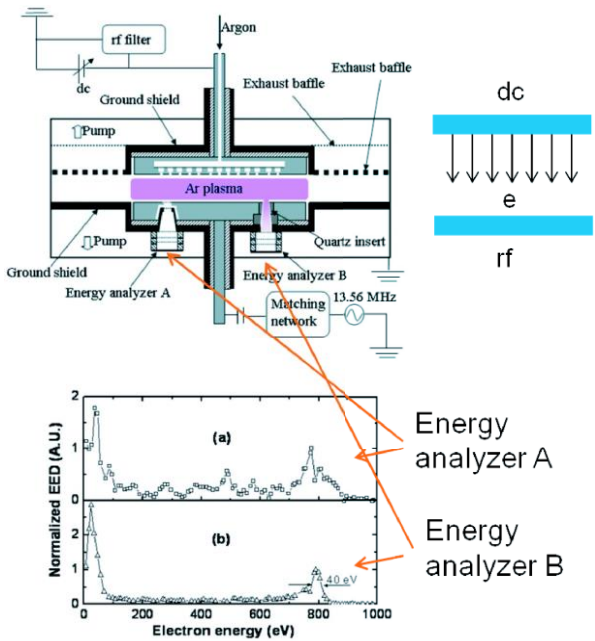

**Fig.1 : A schematic of experimental RF-DC device from Ref. [6], and the electron distributions measured in it. Bulk electrons are sensed by detector B and those that reach the electrode are sensed by detector A.**

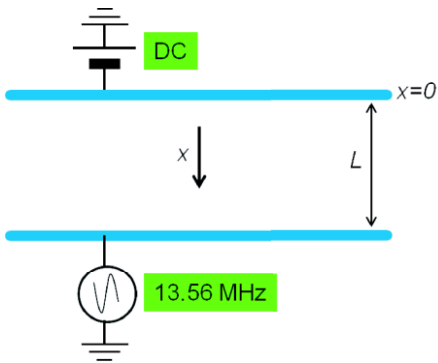

**Fig.2: Schematic diagram of the RF-DC device under study.**

Note that this is a simplified configuration, and plasma sources for etching would typically have a high-frequency source (e.g. 60 MHz) coupled to the DC source for the purpose of generating the plasma at higher density. Two detectors were utilized in the experiment, one (detector "A"), which has a sub-Debye mesh and a restricted aperture for measuring the time-averaged electron energy probability function (EEPF) of the flux impinging upon the RF electrode, and another one (detector "B"), with a super-Debye opening for measuring the distribution in the quasi-neutral plasma bulk by extracting the electrons from a sufficient distance away from the RF sheath region. The experiment operated at 13.56 MHz with 3 cm discharge gap in argon gas at 50 mTorr. Peak-to-peak RF amplitude was 2100 V, and the applied DC bias potential was 800 V. An interesting finding from the measurements with the detector A (distribution at the RF electrode) was that the distribution of electrons impinging upon the RF electrode contains not only the "ballistic" peak at the energy corresponding to the DC bias, but also a structured plateau with multiple peaks, see Fig. 1. It is these peaks in the plateau between the high-energy tail and the thermal group that are believed to provide the beneficial properties of RF-DC plasma in material-processing applications, namely an enhanced fraction of energetic electrons impacting the surface. In other experiments [10], peaks on the EVDF inside the quasi-neutral plasma were also observed. The main goal of the present work is to seek insight into these structures by studying the motion of secondary electrons interacting with DC and RF sheaths, and derive analytic relations connecting the properties of the velocity distribution with the discharge parameters.

## II.1. Numerical model

Our numerical model is a one-dimensional (1D3V) test-particle representation of the discharge device of Ref. [10]. The idealized configuration is shown in Fig. 2. The primary assumption in the model is that the energetic electron group, originated through secondary emission and accelerated by electric field in the electrode sheaths, traverses the plasma with only few large-angle scattering collisions. The model requires specification of the potential profile $\Phi(x,t)$ in the discharge. This is done considering the fact that the area of the ground surface available to collect the high-frequency current is several times larger (~ 5x) than that of the powered electrode. Therefore, the self-bias at the RF electrode is high and the plasma potential (and its oscillation amplitude) with respect to the grounded surface is low.

A parabolic spatial profile of electrostatic potential within the DC and RF sheaths was adopted, corresponding to an assumption of constant ion density and a step-like bulk electron density profiles in the sheath transition. The sheath width in this model is proportional to the square root of the potential. Thus the ion-sheath model retains a nonlinear relationship between voltage and current oscillations. This assumption is not essential in a test-particle problem because the sheaths are collisionless for electrons and details of the potential profile inside the sheath do not affect electron trajectories (the assumed collisionless regime also implies that the flux of electrons produced by impact ionization within the sheaths, and subsequently accelerated, is small compared to the surface-emitted flux). The potential inside the plasma (between the sheaths) was assumed to be uniform; that is a small electric field in the plasma was assumed not to affect the high-energy electrons.

In these 1D simulations of test particle dynamics, it was convenient to choose the plasma potential as the zero reference value. The ion-induced electron emission from the electrodes is accounted for by injecting steady fluxes of electrons at $x = 0$ and $x = L$. Focusing only on emitted electrons, under a test-particle approximation, allowed us to avoid noise and poor statistic issues for particle-in-cell codes and resolve the details of EVDF.

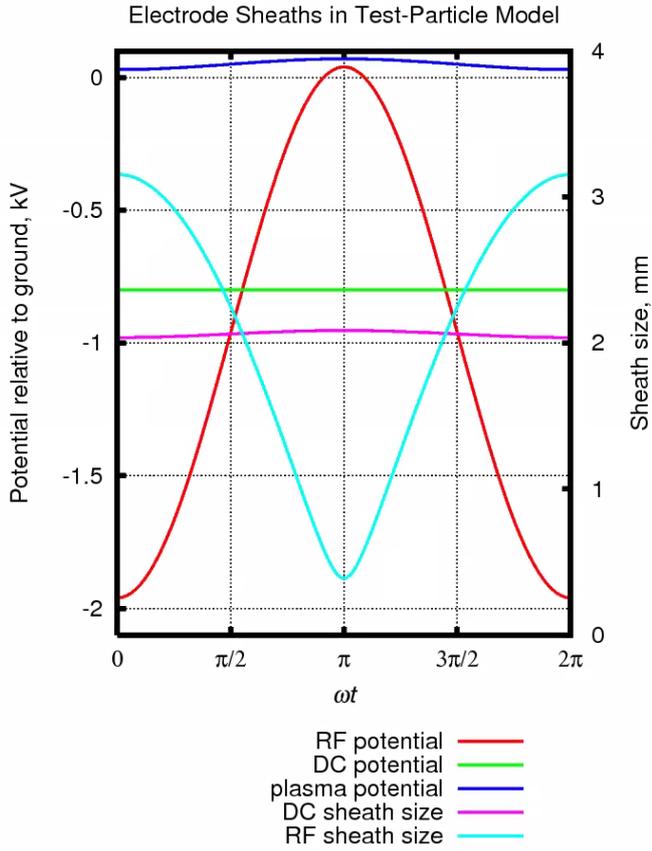

Fig. 1: RF and DC potentials and corresponding sheath widths as functions of time.

While the ion fluxes driving secondary electron emission at either electrode in a non-symmetric system may be different, and the materials facing the plasma on either electrode can be different as well, for simplicity we choose injected electron fluxes to be equal. Electron-electron secondary emission is introduced according to an analytical approximation of the yield curve according to Ref. [13]. The EVDF and the resulting fluxes scale proportionally to the specified injected flux in a steady state. As already noted, potential profiles considered in this paper correspond to a limiting case of the small ratio between the area of the capacitively driven RF electrode and that of the RF grounding surface, and complete filtering out of the RF current in the DC bias circuit. This determines the structure, in space and time, of the electrode sheaths. Under these conditions, the RF electrode potential relative to the plasma can be expressed as:

$$V_{RF}(t) = -\frac{1}{2}(1-\alpha)V_{pp}(1+\cos\omega t) - V_p, \qquad (1)$$

where $V_{pp}$ is the peak-to-peak RF voltage (relative to the grounded wall collecting the RF current), typically in the kV range, and $V_p$ is the minimum voltage of the order of ~10 eV determined by the floating condition with regard to DC current at the RF electrode [14]. The DC electrode voltage relative to the plasma potential reads

$$V_{DC}(t) = -V_{DC}^0 - \alpha\frac{1}{2}V_{pp}(1-\cos\omega t) - V_p, \qquad (2)$$

where $V_{DC}^0$ is the imposed DC bias. The small parameter $\alpha$, controlling the ratio of oscillation amplitudes at the RF ground and at the powered electrode is introduced to account for the small but finite area ratio $A_{RF}/A_\text{ground}$ between the respective surfaces. It is known to be proportional to $(A_{RF}/A_\text{ground})^q$, where $q \geq 2$ [14]. In the simulations reported here, we set $\alpha = 1/50$ (although, for high electron energies of interest to us, an idealized case $\alpha = 0$ was sufficient to obtain the results). The value of $V_p$ was set at 30 Volt. An example of a waveform over one RF cycle, with $V_{pp} = 2000\ V$ and $V_{DC}^0 = 800\ V$, is shown in Fig. 1. These values are within the range of typical experimental conditions. The sizes of electrode sheaths, in accordance with the approximation of constant ion density, were calculated in proportion to the square root of the potential difference, that is $l_{RF,DC}(t) = l_0 \sqrt{V_{RF,DC}(t)/V_{DC}^0}$, where $l_0 = 0.2$ cm. Electrons are injected at a constant rate from both electrodes and their motion is simulated over multiple RF cycles until a steady-state EVDF is formed. The primary output of the calculations is the EVDF averaged over a full RF cycle. The flux rate and the energy grid for EVDF collection were chosen in order to resolve the structure predicted by the analysis, with a sufficient number of particles (≈250 in the present case) collected into each bin over the RF cycle to control statistical noise. The practically adopted energy grid was non-uniform, with $\Delta E = 1.2\text{ eV} + 4 \times 10^{-3} E$ to improve counting statistics at higher energies where the distribution falls off, while keeping the resolution at a sufficient level to observe the expected structures.

## III. Electron Energy Distribution Function in the Collisionless Limit

Results of simulations for typical experimental parameters of Refs. [6], [10], but with electron-neutral collisions turned off, are shown in Fig. 2. The discharge gap is 3 cm, and the driving frequency is 13.56 MHz. The DC bias voltage is 800V and the RF peak-to-peak amplitudes are 1500, 2000, 3000 V in cases a, b, and c, respectively. The plots represent cycle-averaged EVDFs of electrons impinging on the surface of the RF electrode. The green line shows the contribution of electrons emitted from the DC electrode, and the blue line shows contribution of electrons emitted from the RF electrode. The red line is the sum of the two.

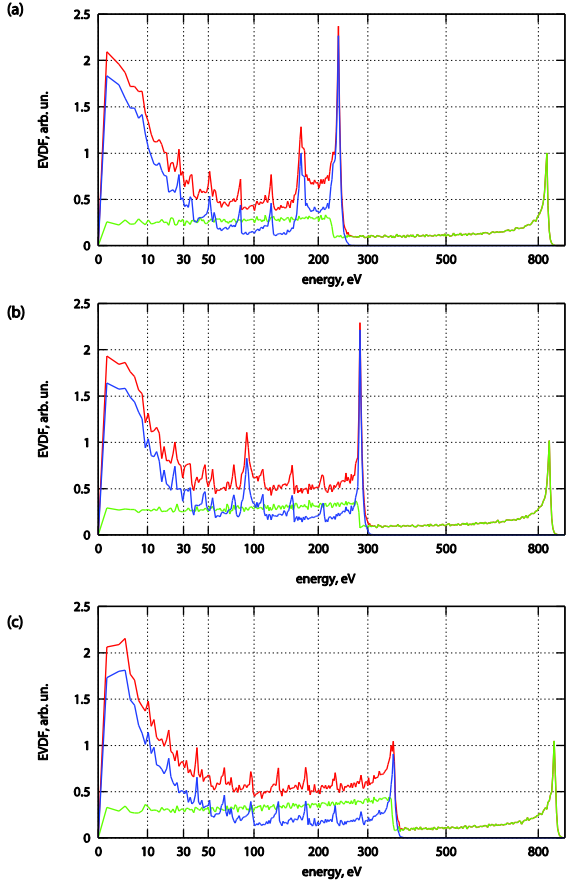

Fig. 2: Electron Velocity Distribution Function at the RF electrode, averaged over the RF cycle, as would have been seen by a detector for $V_{DC}^0 = 800V$ and: (a) $V_{pp} = 1500V$; (b) $V_{pp} = 2000V$; (c) $V_{pp} = 3000V$. The green line shows the contribution of electrons emitted from the DC electrode and the blue line shows contribution of electrons emitted from the RF electrode. The red line is the sum of the two. The linear velocity scale is labeled in energy units.

It is evident from Fig. 2 that the distributions of electrons originating from the DC electrode and of those from the RF electrode are quite different. The EVDF of electrons emitted from the DC electrode exhibits a pronounced peak at the energy near the DC bias and a raised plateau below a certain threshold $E_{thr}$. The EVDF of electrons emitted from the RF electrode exhibits a series of pronounced peaks in the range of energies close to that of the plateau. In order to observe these structures, the simulation required a good energy resolution, as noted earlier. A raised plateau in the EVDF can be seen in the results reported in Ref. [15] where bias was applied to a symmetric discharge (the present authors believe that, at high bias, those results can be interpreted in the same manner as in this work). However, multiple peaks were not observed in previous simulations, because those features may cease to exist in presence of collisions, as will be discussed in Sect. IV.

The step-and-peaks structure of EVDF is quite sensitive to the shape and amplitude of the sheath potential waveforms and can be viewed as a tool for their diagnostics. These properties are difficult to measure directly without distorting the experimental results. Similar findings have already been made regarding the distributions of ions. For example, Israel *et al.* [16] demonstrated how locations of peaks in ion distribution functions (IDF) can be utilized to evaluate the cross-section of charge-exchange collisions. O'Connell *et al.* [17] showed that measured ion distributions carry information about both the width and the potential amplitude in the sheath, as well as about how strongly the waveform deviates from a sinusoid. Thus, in general, analytical and numerical results combined with experimental data for IDF or EVDF may provide a lot of detailed information about sheath structure that is not possible to assess with in-situ measurements. Having this in mind, we developed a theoretical analysis that connects EVDF structure to the sheath voltage waveform. We describe it in the following sections.

### III.1. EVDF of electrons originating at the DC electrode

First, we consider the electrons emitted from the DC electrode and exiting at the RF electrode. The electron trajectory depends on the phase of the RF waveform at the moment of emission. Let $\Phi_A$ be the phase at which the RF sheath potential becomes equal to the sheath potential at the biased electrode. Ignoring the potential (both constant and oscillating) in the plasma relative to the ground, that value is defined simply by

$$V_{RF}(\Phi_A) = V_{DC}. \tag{3}$$

When the voltage at the DC electrode exceeds the RF voltage, electrons exit to the RF electrode in a single pass through the gap. For a harmonic RF waveform,

$$V_{RF}(\phi) = \frac{V_{pp}}{2}(1 + \cos \phi), \tag{4}$$

the phase interval within which electrons traverse the discharge gap in *a single pass* is given by

$$\Phi_A < \phi < 2\pi - \Phi_A, \tag{5}$$

as demonstrated in Fig. 5, where

$$\Phi_A = \cos^{-1}\left(\frac{2V_{DC}}{V_{pp}} - 1\right). \tag{6}$$

During the phase interval

$$-\Phi_A < \phi < 0, \tag{7}$$

the RF voltage exceeds the DC voltage; electrons emitted from the DC electrode are reflected from the RF sheath and gain some energy because the sheath expands into the plasma. Therefore, they overcome the DC potential barrier and exit the plasma at the DC electrode side after performing a double-pass in the discharge gap. Hence we call this phase interval a *double-pass interval*.

Next, electrons emitted from the DC electrode during the phase interval given by

$$0 < \phi < \Phi_A \tag{8}$$

also bounce back off the RF sheath, but they lose some energy in collisions with the receding sheath and hence become trapped until the RF sheath potential becomes smaller than the DC bias. Hence, we call this time period a *trapping interval,* see Fig. 3. All electrons accumulated during this phase are dumped during the RF phase interval

$$\Phi_A < \phi < \Phi_B, \tag{9}$$

where the difference $\Phi_B - \Phi_A$ equals $\omega\tau$, $\tau$ being the bounce period of electrons traveling from the DC electrode to the RF sheath and back. Indeed, an electron that collides with the RF sheath at the phase slightly earlier than $\Phi_A$, but past $(\Phi_A - \omega\tau)$, would traverse the discharge gap again, back and forth, and then exit to the RF electrode. This portion of the RF cycle can be called the "dumping interval".

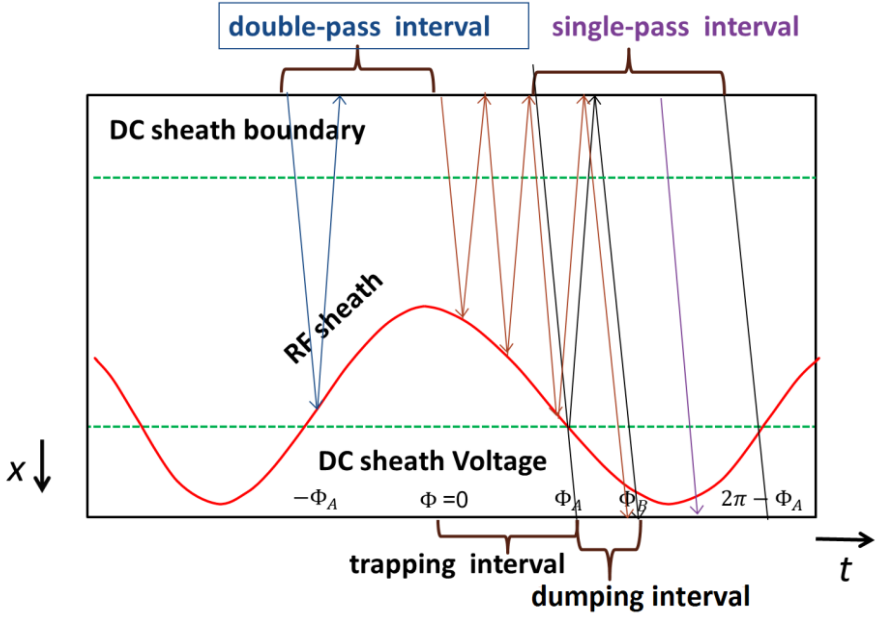

Fig. 3: Schematics of electron trajectories in RF-DC discharge in (t, x) phase space. RF electrode is positioned at the bottom and DC at the top. Red curve shows schematically the position of the RF sheath boundary, the green curve - the position of the DC sheath boundary. The trapping, dumping and single-pass intervals show RF phases with different dynamics for electrons emitted from the DC electrode.

Calculation of the EVDF observed at the RF electrode is straightforward for the single-pass interval, Eq. (5). Assuming a constant flux of emitted electrons, $\Gamma_0$, the EVDF at the edge of the DC sheath reads (with energy value as the function argument):

$$f_0(E) = m\Gamma_0 \delta(E - V_{DC}).$$

The normalization is $\int_0^{+\infty} f(v)dv = n_o = \Gamma_0/v_0$. The energy of electrons reaching the RF electrode is given by

$$E(\phi) = V_{DC} - V_{RF}(\phi), \qquad (10)$$

so that the time-dependent EVDF at the RF electrode is

$$f(E, \phi) = f_0(E + V_{RF}(\phi)) = m\Gamma_0 \delta[E - E(\phi)]; \ E(\phi) > 0. \qquad (11)$$

The time-averaged EVDF of single-pass electrons escaping to the RF electrode is obtained by averaging over the RF cycle:

$$<f(E)> = \frac{1}{2\pi}\int_0^{2\pi} f(E, \phi)d\phi. \qquad (12)$$

With $f(E, \phi)$ from Eq.(11) and $E(\phi)$ from Eq.(10), the integration gives

$$<f(E)> = \frac{m\Gamma_0}{2\pi}\int_0^{2\pi} d\phi \delta[E - V_{DC} + V_{RF}(\phi)] = \frac{1}{\pi}\frac{m\Gamma_0}{\left|\frac{dV_{RF}}{d\phi}\right|_{V_{DC}-V(\phi)=E}}. \qquad (13)$$

Here, it was taken into account that there are two values of $\phi$, symmetric about $\pi$, satisfying Eq.(10). Substituting the voltage derivative as a function of electron energy at the RF electrode

$$|dV_{RF}/d\phi| = \frac{1}{2}V_{pp}|\sin \phi|$$
$$= \sqrt{1/2\, V_{pp}(1+\cos\phi)\, 1/2\, V_{pp}(1-\cos\phi)} \quad , \quad (14)$$

and then utilizing Eq.(10) yields

$$<f_{pass}(E)> = \frac{m\Gamma_0}{2\pi}\sqrt{\frac{V_{pp}}{(V_{DC}-E)(V_{pp}-V_{DC}+E)}}. \quad (15)$$

The singularity at $E = V_{DC}$ occurs because $dV_{RF}/d\phi = 0$ when $V_{RF} = 0$ at $\phi = \pi + 2\pi n$. In plasma physics, a distribution of this type, for ions traversing a low-frequency sheath, was found at least as early as in 1967 [18], after being discovered in vacuum electronics.

Now let us consider the distribution of the electrons lost to the RF electrode during the "dumping interval" identified by Eq.(9). All electrons emitted within the trapping interval $0 < \phi < \Phi_A$ will exit to the RF electrode during the dumping interval $\Phi_A < \phi < \Phi_B$, as shown in Fig. 3. The trapping interval is determined by the ratio of the RF amplitude to the bias potential and is typically a fraction of the RF period. The dumping interval is determined by the time it takes an electron to bounce back and forth between electrodes; $\Phi_B - \Phi_A = 2\omega L/v_0 \ll 1$, where $\omega = 2\pi f$ is the discharge frequency; $v_0 = \sqrt{2eV_{DC}/m}$ is the velocity of an electron emitted from the DC electrode and acquiring kinetic energy $eV_{DC}$; and $L$ is the electrode spacing. Due to this accumulation effect, the EVDF contribution from the trapping interval is $N$ times higher than the single-pass contribution, where

$$N = \frac{\Phi_A}{2(\Phi_B - \Phi_A)} \approx \frac{\cos^{-1}\left(\frac{2V_{DC}}{V_{pp}}-1\right)\sqrt{2eV_{DC}/m}}{4\omega L}. \quad (16)$$

The factor of 1/2 is again due to the two branches ($\Phi_A < \phi < \pi, \pi < \phi < 2\pi - \Phi_A$) of the pass-through distribution. Therefore, the contribution into the EVDF during the dumping period is the value given by Eq.(15), times $N$, within the energy range $E < E_1$, where $E_1$ is energy gained by an electron upon exiting near the phase $\Phi_B$, see Fig. 3:

$$E_1 \approx \frac{2\omega L}{v_0}e\left|\frac{dV}{d\phi}\right|_{\phi=\Phi_A} = \omega L\sqrt{2em(V_{pp}-V_{DC})}. \quad (17)$$

For example, for the conditions of Fig. 2, the width of the step-like structure calculated from (17) gives $E_1$ =227 eV, 297 eV, 402 eV, and the relative jump in height calculated from (16) is $N$ = 2.48, 2.92, 3.39 for $V_{pp}$= 1500, 2000, and 3000 V, respectively. These predictions agree well with the properties of the step-like structures (in the green curves) on Fig. 2. There will be, in fact, an additional small step-down at some value below $E_1$, except in special cases when the trapping interval $\Phi_A$ is an integer multiple of the bouncing phase $\omega\tau$. We will not dwell upon this detail. We also note that the above analysis should apply to some of the step-like distributions apparently present in numerical results of Diomede *et al.* [15], particularly at high values of the bias voltage (applied in that case to a symmetric discharge).

Note that we neglected a small change in kinetic energy that occurs in each interaction of a bouncing electron with a moving sheath. For energetic electrons with $\omega\tau/2\pi \ll 1$ we are interested in, such energy change is bounded, due to the conservation of adiabatic invariant, and therefore the relative change of electron energy is small, on the order of $L_{sh}/L$; see [19] for more details.

### III.2. EVDF of electrons emitted from, and returning to the RF electrode

The EVDF of electrons emitted from the RF electrode can be treated similarly to that of electrons originating from the DC electrode, as illustrated in Fig. 4. The new effect is that the bouncing time is a function of the RF phase.

During the phase interval

$$-\Phi_A < \phi < \Phi_A \tag{18}$$

the RF sheath potential exceeds the DC sheath potential; electrons emitted from the RF electrode exit at the DC electrode and are not reflected back towards the RF electrode. Because we are primarily interested in the EVDF at the wafer placed onto the powered electrode, electrons emitted during this portion of the cycle will be disregarded.

Electrons emitted from the RF electrode during the phase interval when the RF sheath voltage is below the DC bias, and decreasing,

$$\Phi_A < \phi < \Phi_C, \tag{19}$$

exit the plasma at the RF electrode side after performing a double-pass (one bounce) in the discharge gap, as shown in Fig. 4. Hence we call this phase interval a *double-pass interval*. The upper boundary of that interval, $\Phi_C$, satisfies the condition $V_{RF}(\Phi_C) = V_{RF}(\Phi_C + \omega\tau(\Phi_C))$, where $\tau(\phi)$ is the bounce time. For a waveform symmetric about the minimum at $\phi = \pi$, the condition for determining $\Phi_C$ becomes

$$\pi - \Phi_C = \omega\tau(\Phi_C)/2 = \omega L/\sqrt{2eV_{RF}(\Phi_C)/m} \tag{20}$$

Note that as $\phi \to \Phi_C$, the energy of escaping electrons approaches zero. Electrons emitted from the RF electrode during the rest of the cycle interval for which the RF sheath potential is below the DC bias, namely

$$\Phi_C < \phi < 2\pi - \Phi_A, \tag{21}$$

also bounce off the DC sheath. Then they bounce off the expanding RF sheath, and keep bouncing. Each of these electrons will remain trapped until it enters the RF sheath in such phase that the potential barrier is smaller than the accelerating potential at the moment the electron was emitted. Hence, we call this time period the *trapping interval*. The electrons accumulated in that phase have a chance to escape during the RF phase intervals

$$\Phi_A + 2\pi n < \phi < 2\pi(n+1) - \Phi_A \tag{22}$$

The periodicity has now been made explicit, because not all of the bouncing electrons will exit within one RF cycle after they were emitted in phases given by (21). The latter is only true for electrons accelerated to energies higher than $V_{RF}(\Phi_C)$, where $\Phi_C$ satisfies Eq.(20), that is, for electrons emitted from the RF electrode at

$$2\pi - \Phi_C < \phi < 2\pi - \Phi_A.$$

On the other hand, if the initial phase satisfies $\Phi_C < \phi < 2\pi - \Phi_C$ (the other, leading portion of the trapping interval defined by Eq.(21)), then the bouncing time $\tau$ is sufficiently large for the RF sheath potential to restore, and become increasing, while the electron travels to the DC sheath and back. These slower electrons can remain in the discharge arbitrarily long (assuming a perfectly coherent waveform). A brief discussion of this property will be given further below.

Let us proceed with the quantitative description. For a secondary electron emitted from the RF electrode at phase $\psi$ and exiting at phase $\phi = \psi + \omega\tau(\psi)$, the energy upon exit to the RF electrode is given by

$$E_{RF}(\phi) = V_{RF}(\psi) - V_{RF}(\phi), \tag{23}$$

and the EVDF at the RF electrode reads, similar to Eq.(11),

$$f_{RF}(E, \phi) = m\Gamma_0 \delta[E - E_{RF}(\phi)]. \tag{24}$$

The time-averaged EVDF at the RF electrode is obtained by averaging over the RF cycle similarly to Eq.(12):

$$<f_{RF}(E)> = \frac{1}{2\pi}\int_0^{2\pi} f_{RF}(E, \phi)d\phi. \tag{25}$$

The important difference with the electrons originated at DC electrode is that the function $E_{RF}(\phi)$ will possess strong dependence on the initial phase, $\psi$. Substituting Eqs.(23) and (24) into Eq.(25) gives

$$<f_{RF}(E)> = \frac{m\Gamma_0}{2\pi}\int_0^{2\pi} d\psi \frac{d\phi}{d\psi}\delta[E - V_{RF}(\psi) + V_{RF}(\phi)] =$$

$$\frac{m\Gamma_0}{2\pi}\left(\frac{1 + \frac{\omega d\tau}{d\psi}}{\left|\frac{dE}{d\psi}\right|}\right)_{E=E_{RF}(\psi)}. \tag{26}$$

Here, we express all quantities as a functions of the initial phase, $\psi$; and

$$\tau \equiv \frac{\phi - \psi}{\omega}$$

is the time elapsed between the emission and the exit of the electron to the RF electrode surface.

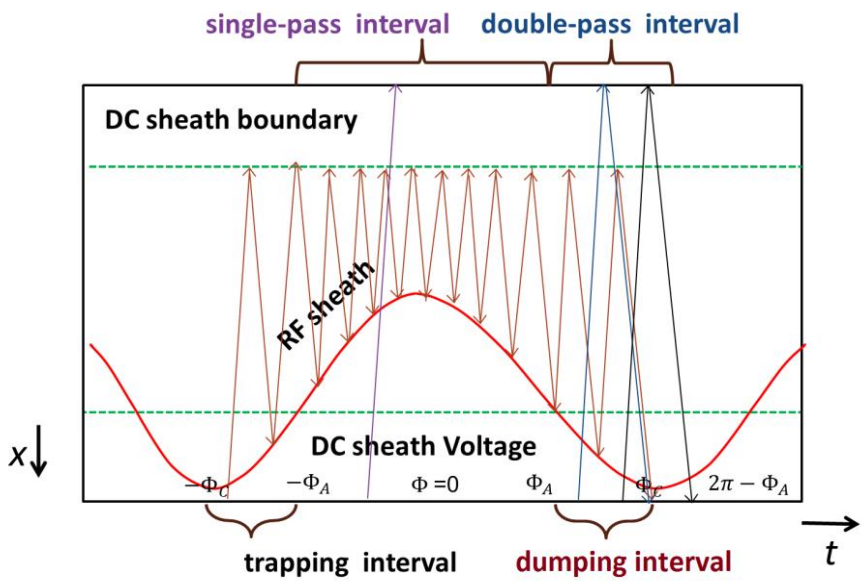

**Fig. 4:** Schematics of electron trajectories in RF-DC discharge in (t, x) phase space. RF electrode is positioned at the bottom and DC at the top. Red curve shows schematically the position of the RF sheath boundary, the green curve - the position of the DC sheath boundary. The trapping, dumping and single-pass intervals show RF phases with very different electron dynamics for electrons emitted from the RF electrode.

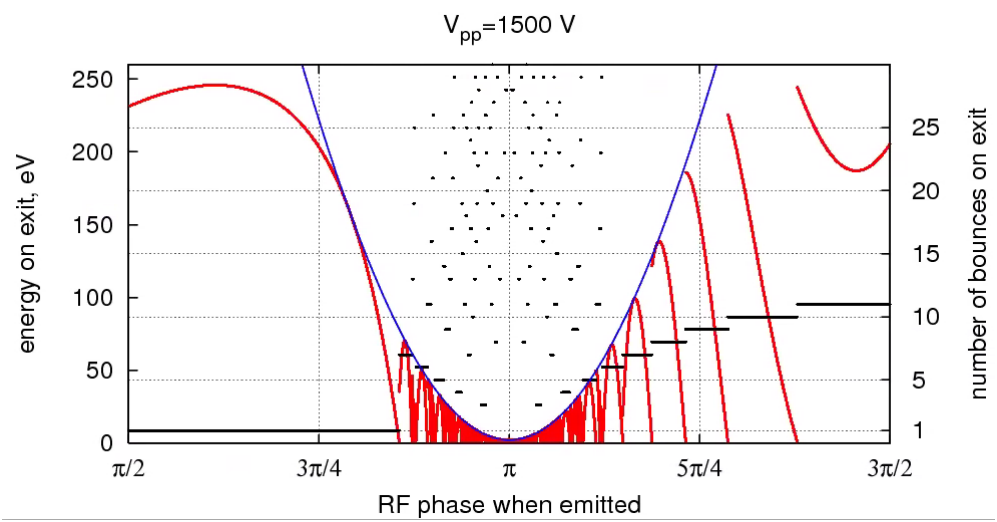

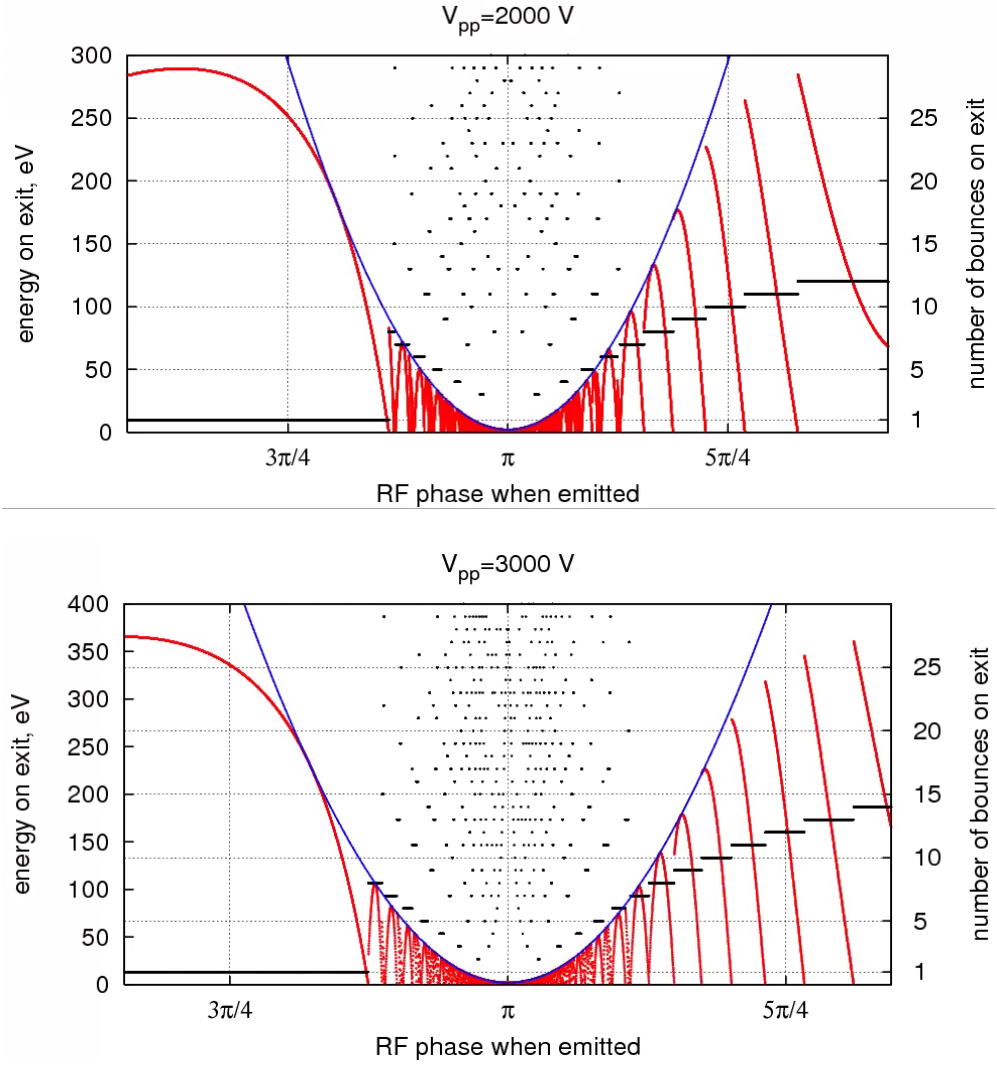

Fig. 5: Plot of energy $E_{RF}(\psi) = V_{RF}(\psi) - V_{RF}(\phi(\psi))$ for electrons originated from the RF electrode at phase $\psi$ and exiting at phase $\phi$. In this simplified calculation the RF voltage minimum is exactly zero and the plasma potential does not oscillate. Right axis shows number of bounces in the gap performed until exiting to the electrode. The blue envelope is $V_{RF}(\psi) - V_{RF}(\pi)$, which is the upper bound for $E_{RF}(\psi)$, attained when $\phi(\psi) = \pi$. The energy of bouncing motion is $V_{RF}(\psi) + 2$ eV.

The function $E_{RF}(\psi)$ is shown in Fig. 5. As evident from this figure, this function is discontinuous and generally non-monotonic within each interval of continuity. Let us first consider phases that correspond to electron trajectories with only one bounce in the gap – so called "double pass interval" given by Eq.(19).

### III.2.1 EEDF of electrons emitted from, and returning to the RF electrode in single reflection from the DC sheath (double-pass interval)

For this interval,

$$E(\psi) = V_{RF}(\psi) - V_{RF}(\phi) = V_{RF}(\psi) - V_{RF}[\psi + \omega\tau(\psi)], \quad (27)$$

$$\tau(\psi) = 2L\sqrt{\frac{m}{2eV_{RF}(\psi)}}.$$

The function $E(\psi)$ has a maximum, because as the voltage decreases, the bounce time increases, which results in increase in $E(\psi)$; eventually, when the emission phase is near $\Phi_C$ given by Eq.(20), the electron exits to the RF electrode with negligible energy. Therefore, the function $E(\psi)$ must have a maximum as evident in Fig. 5.

To determine the maximum of the function $E(\psi)$ analytically we can use a small parameter

$$\epsilon \equiv \omega\tau(0) = \omega L \sqrt{\frac{2m}{eV_{PP}}}, \qquad (28)$$

which is twice the one-way travel time, in phase units, of an electron emitted at the maximum of the RF voltage phase $V_{PP} = V_{RF}(0)$ (such electrons obviously escape to the DC electrode). Assuming for generality that the RF waveform can be scaled with a single parameter – its maximum value,

$$V_{RF}(\psi) \equiv V_{PP} g(\psi),$$

Eq. (27) can be rewritten in the form

$$E(\psi)/V_{PP} = g(\psi) - g\left[\psi + \epsilon/\sqrt{g(\psi)}\right]. \qquad (29).$$

Simple linearization of Eq.(29) will not be sufficient to locate the minimum of $E(\psi)$ in the seemingly simple case of a sinusoidal waveform,

$$g(\psi) = (1 + \cos\psi)/2. \qquad (30)$$

Specifically, linearization of Eq. (29) yields

$$E(\psi) \approx -\epsilon V_{pp} g^{-\frac{1}{2}} \frac{dg}{d\psi} = -2\epsilon \frac{d}{d\psi}\sqrt{g(\psi)}. \qquad (31)$$

Then for $g(\psi)$ from Eq.(30), $E(\psi) \approx \epsilon |\sin(\psi/2)|$, which has a maximum at $\psi = \pi$ where $\tau = \infty$ and the linear approximation is not valid. Therefore, the correct approach is to expand to the second order in $\epsilon/\sqrt{g}$:

$$E(\psi) \approx -2\epsilon \frac{d}{d\psi}\sqrt{g(\psi)} - \frac{1}{2}\epsilon^2 \frac{g''}{g}. \qquad (32)$$

For a harmonic waveform $g(\psi)$ given by Eq.(30), the above becomes

$$E(\psi) = \epsilon \sin\frac{\psi}{2} + \frac{1}{2}\epsilon^2 \frac{\cos\psi}{1+\cos\psi}.$$

Solving $dE/d\psi=0$ gives $\psi_*$, the value of the initial phase maximizing the energy of the single-bouncing electrons upon exiting through the RF sheath. For a sinusoidal waveform, the value $\psi_*$ can be found without further approximations. We present the result in powers of $\epsilon$, as can be done for any other waveform. The maximum of $E(\psi)$ occurs at $\psi_* = \pi - (8\epsilon)^{1/4} - \frac{1}{96}(8\epsilon)^{3/4} + \mathcal{O}(\epsilon^{5/4})$. For example, for a practically relevant value $\epsilon = 0.2$, we have $(\pi - \psi_*) \approx 1.14$ rad $\approx 65°$. It should be noted that dependence of $(\pi - \psi_*)$ upon RF amplitude is weak, namely $V_{pp}^{-1/8}$. The corresponding maximum in energy is

$$E_* = V_{pp}\left(\epsilon - \frac{1}{\sqrt{2}}\epsilon^{3/2} + \frac{7}{16}\epsilon^2 + \mathcal{O}(\epsilon^{5/2})\right). \tag{33}$$

It scales as $V_{pp}^{1/2}$. Table 1 shows good agreement between simulations and Eq.(34). The phase value for which $E(\psi) = 0$ also can be found. To the lowest order in $\epsilon$, it equals $\pi - \sqrt{\epsilon}$, which is, of course, $\Phi_C$ defined by Eq.(20). Also note that the other, lower, boundary for the initial phase of the single-bouncing group is $\psi = \Phi_A$ (at which $V_{RF}(\psi) = V_{DC}$, Eq. (18)), and the corresponding energy value $E(\psi)$ is the same as that for the step in the EVDF of electrons emitted from the DC electrode: it is $E_1$ given by Eq. (17), which can also be written in terms of $\epsilon$ as $E_1 = \epsilon\sqrt{V_{pp}(V_{pp} - V_{DC})}$.

Table 1: The maximum $E_*$ according to Eq.(34) and to numerical results shown in Fig. 5.

| $V_{pp}$ [V] | $\epsilon$ | $\psi_*$ [rad] | $E_*$ Eq.(34) [eV] | $E_*$ from Fig. 5 [eV] |
|---|---|---|---|---|
| 1500 | 0.221 | 1.97 | 253 | 247 |
| 2000 | 0.192 | 2.01 | 297 | 291 |
| 3000 | 0.156 | 2.07 | 369 | 365 |

The maximum of the function $E(\psi)$ defines a location of a peak in the EVDF, in accordance with Eq.(26). The energy value where the peak occurs, given by Eq.(33), does not depend upon the DC bias but is seen to be close to $E_1$ under the condition $V_{pp} \geq 2V_0$. This explains why the step in the distribution of the DC-emitted electrons and the peak of the single-reflected electrons originating from the RF electrode occur near each other in the simulated EVDF, as evident from Fig. 2. It should also be noted (as seen on the same figure) that the maximum location $\psi_* \approx \pi - (8\epsilon)^{1/4}$ is much closer to $\Phi_A$ than it is to $\Phi_C \approx \pi - \epsilon^{1/2}$.

The important observation is that the structure of the EVDF of the secondary electrons reflected back to the RF electrode is sensitive to the smoothness of the voltage waveform, as indicated by the appearance of higher derivatives of the driving voltage.   Indeed, it will be seen that the shape of the distribution is strongly affected by the presence of high harmonics even at a small amplitude relative to the fundamental, as further explained in Sect. V.

### III.2.2 EVDF of electrons emitted from, and returning to the RF electrode after multiple bounces off the DC sheath

According to Eq.(26), each stationary point of the function $E(\psi)$ defines a peak in the EVDF at that energy value.   As is evident from Fig. 5, *multiple peaks appear due to the presence of several groups of electrons, each undergoing a different number of bounces between the two sheaths prior to escaping.*

For each such group (theoretically, an infinite number; to be discussed below), there could be one or more maxima and/or minima of exit energy vs. starting phase. To illustrate the formation of multiple peaks, we utilize Fig. 6, Fig. 7, and Fig. 8. These figures visualize simulations for three cases, differing by the amplitude of the driving RF voltage.

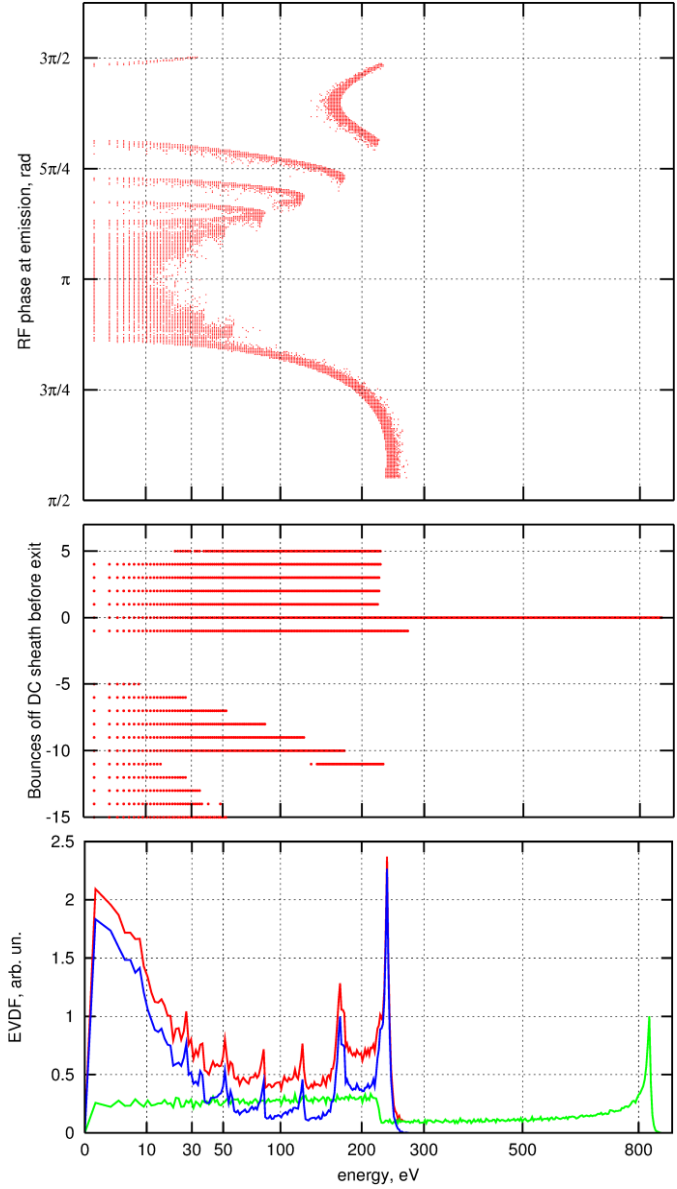

Fig. 6 Top: Scatter plot in $(E, \psi)$ plane of the starting phase $\psi$ and the exit energy $E_{RF}(\psi) = V_{RF}(\psi) - V_{RF}(\phi(\psi))$ for electrons emitted from the RF electrode and escaping to it at phase $\phi(\psi)$. The abscissa is velocity, converted to eV. Electrons emitted from the DC electrode are not shown in this panel; their final energy follows the RF waveform. Middle: Numbers of bounces electrons perform in the gap before exiting to the RF electrode. Non-negative numbers are for electrons emitted from the DC electrode, negative numbers are for electrons originating from the powered electrode. Bottom: The EVDF of electrons at the RF electrode averaged over RF period as would be seen by an energy analyzer for $V_{DC} = 800V$ and $V_{pp} = 1500V$. The green line shows contribution of electrons emitted from the DC electrode and blue line shows contribution of electrons emitted from the RF electrode. The red line is the sum of the two.

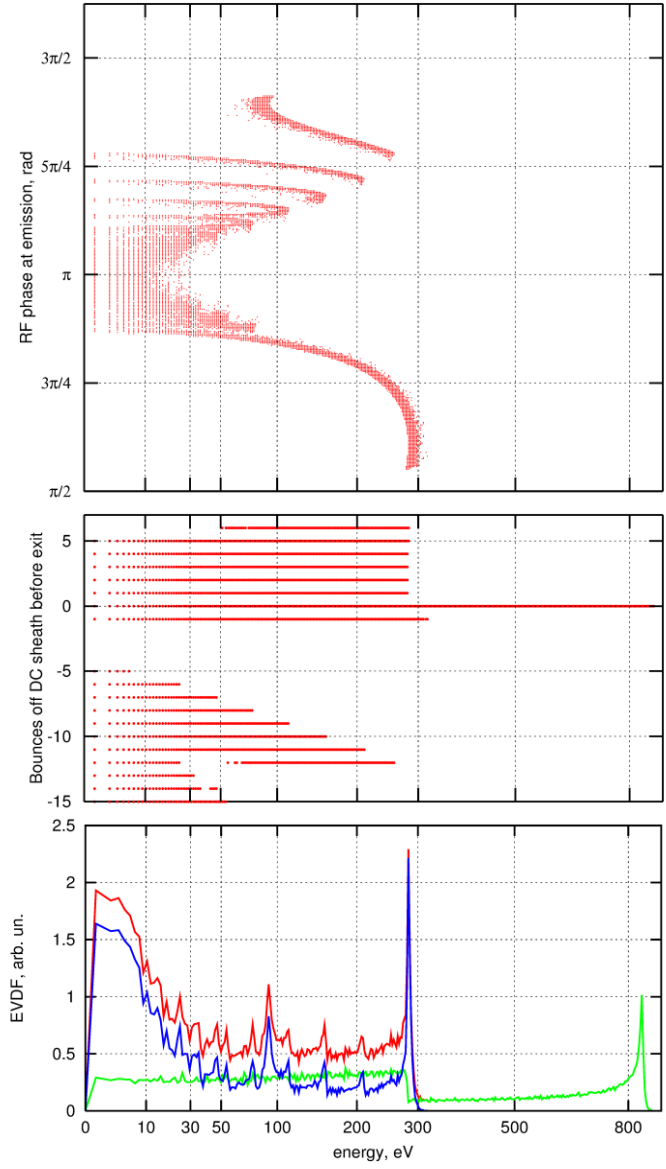

**Fig. 7** Same as Fig. 6 but $V_{pp} = 2000V$.

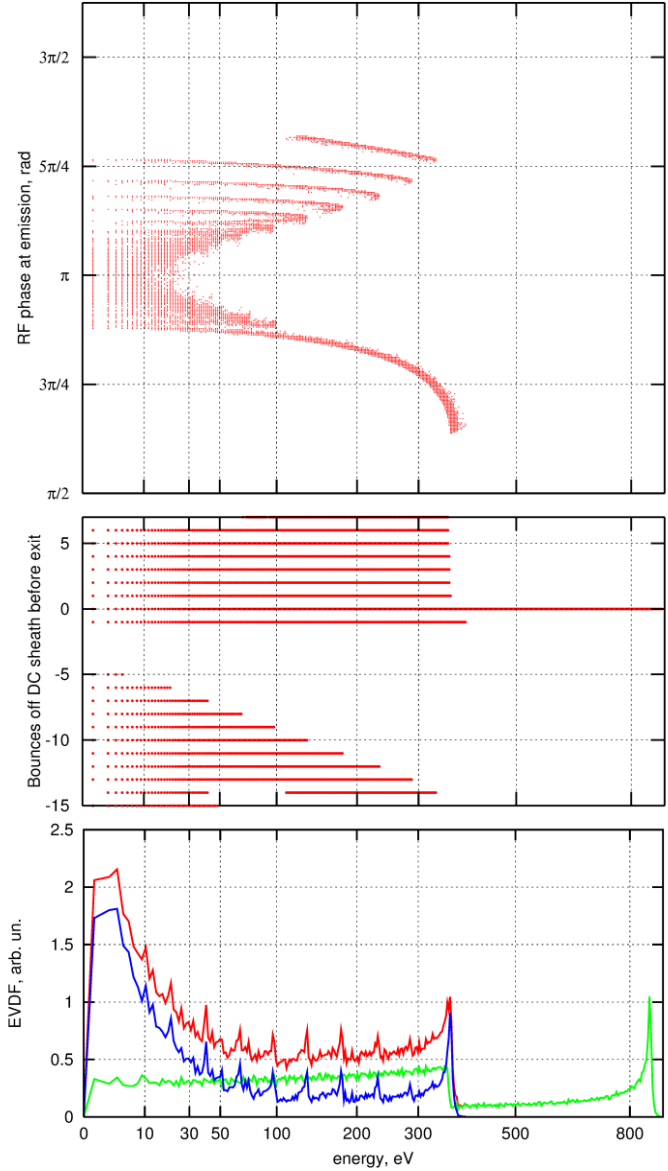

Fig. 8: Same as Fig. 6 but $V_{pp} = 3000V$.

The maxima and minima of $E(\psi)$ for all possible trajectories are shown to give the locations of peaks in the numerically generated EVDF. The top panel for each case is a scatter plot of the inverse function to $E(\psi)$, obtained by following the electrons which escape to the RF electrode. Only the electrons originating from the RF electrode itself are counted in this plot. For the electrons originating at the DC electrode, the picture is much simpler and has been fully discussed in Sect. III.1. Additional visualization of how the EVDF is formed is offered by plotting the number of bounces before exiting to the RF electrode versus detected electron energy at the RF electrode. Such plot is presented in the middle panel. It is simply a series of discrete intervals, each representing the energy range for a group which bounces a given number of times. For clarity, the density of the dots in this case does not represent the magnitude of the EVDF, but is set to be constant where the electrons are present, and zero elsewhere. A *negative sign* has been assigned to the bounces underwent by the electrons *originating from the RF electrode* and the *positive sign* has been assigned to the bounces underwent by the electrons *originating from the DC electrode* so they do not overlap on the diagram. Zero bounce count is reserved for the ballistic electrons traveling directly from the DC electrode to the RF electrode during the single pass-through interval given by Eq.(5), see Fig. 3.

This plot clearly shows that the step-like structure in the EVDF is due to the temporarily trapped population of the DC-emitted electrons. For example, as evident from Fig. 6 for $V_{pp} = 1500V$, there are 5 groups of such electrons, bouncing back of the DC sheath from 1 to 5 times before exiting to the RF electrode. The energy range of the group with the highest number of bounces does not extend down to zero. This is so because the trapping interval is not an exact multiple of the bouncing time. The maximum value, on the other hand, is the same for each of those groups.

For electrons emitted from the RF electrode and accelerated in the RF sheath when $V_{RF} < V_{DC}$, the number of bounces can be high. The scale on Fig.8 – Fig.10 is cut off at 15 bounces, but in the idealized model that number can be arbitrarily large (infinite, if electrons are emitted with zero energy and the minimal value of self-biased RF signal is zero). In order for any electron to escape back to the RF electrode surface upon bouncing off the DC sheath, the phase $\phi$ of its return to the RF sheath boundary has to fall within the escape window $\psi < \phi < 2\pi - \psi$ (mod $2\pi$), where $\psi$ is the phase of emission. Let us consider electrons with starting phases $\psi$ in the interval $\Phi_C < \psi < 2\pi - \Phi_C$, where $\Phi_C$ is defined by Eq.(20). The bouncing time (in units of RF phase) for those electrons will be larger than the escape window. Thus, it is possible for a bouncing electron to "leap over" the escape window, in which case it would remain trapped for [at least] one more RF cycle. Correspondingly, in the collisionless model, there is a concentration of EVDF peaks at low energies, due to electrons emitted close to the minimum of the driving voltage. For an electron emitted from the RF electrode at phase $\pi + \Delta\psi$ near the minimum of the RF sheath potential, the condition of returning to the electrode is:

$$-|\Delta\psi| \leq \Delta\psi + n\tau - mT \leq |\Delta\psi|, \tag{34}$$

where $\tau = \tau(\psi)$, again, is the bounce period, $T \equiv 2\pi/\omega$ is the RF period, and $(n, m)$ is a pair of positive integers for which the inequality is satisfied with the smallest possible $n$. Eq.(35) makes apparent the interesting properties of the trapped population, such as, for example, its fractal nature in the phase space.

Fig. 11 illustrates the evolution of electrons emitted from the RF electrode with the aid of a Poincare map. The radius in the plot is the constant velocity of the emitted electron, acquired in the initial pass through RF sheath. The polar angle is the RF phase at the moment of each interaction with the RF sheath (from emission, through each return after bouncing off the DC sheath, to exiting back to the electrode). The polar radius is the magnitude of the velocity, $v$ normalized by its value $v_{trap}$ at the phase $\psi = \Phi_c$, at which moment the bouncing time $\tau$ and the escape window become equal. The void in the center is due to finite initial energy of emitted electrons, 2 eV, chosen to avoid infinite bouncing times. The minimum of the RF signal for the map was exactly zero, unlike in the test-particle runs. Each orbit, traversed counterclockwise, originates from one of the two branches of the grey curve $v = v(\psi)$; $\Phi_A < \psi < 2\pi - \Phi_A$ is the condition for bouncing back off the DC sheath. The map can be viewed as a time-dependent transformation of that curve. It is seen that orbits spanning more than one RF cycle can exist if $v/v_{trap} < 1$ ( $\Phi_c < \psi < 2\pi - \Phi_c$). Fig. 11 shows examples of individual orbits with given numbers of bounces. It aids in understanding the complete map, and also brings to attention the following: at sufficiently low pressures, it may be possible to observe EVDF peaks attributed to groups of electrons with large number of bounces. Indeed, the energy levels of the one-bounce and of the "fast" 11-bounce are seen to overlap (one originates at $\psi < \pi$ and the other at $\psi > \pi$), meaning the 11-bounce peak could be possible to observe if the gas pressure is reduced some 10-fold relative to the collisional cases in this paper, i.e. to the range of few mTorr. The experiment would have to be accurate, because the 11-bounce group in our example is about 10 times weaker in terms of the flux, based on its initial phase interval (portion of the grey curve) vs the one-bounce group. Good coherency of the driving waveform would also be required.

A more detailed analysis would have to account for bounce-to-bounce variation in $\tau$ due to interaction with the moving RF sheath. It would be, essentially, another take on the collisionless heating, because of non-adiabatic motion of the slow electrons. It will not be pursued here. The map shown in Figs. 11-13 was generated with no time variation in the bouncing period, to focus on the effect we have been discussing.

The value of the collisionless limit is that it shows how complex the EVDF structure could be even in the simplest case of RF-DC discharge in test-particle approximation. At the same time, applicability of the model is limited to very low pressures because it is assumed that electrons performing multiple bounces do not suffer collisions. The effect of scattering will be examined in the next section. Another insight drawn from the above analysis is sensitive dependence of the measured EVDF upon the waveform of the driving voltage. This leads to a prediction that more structures should arise due to presence of higher-order harmonics. Such effect will be examined in the final section.

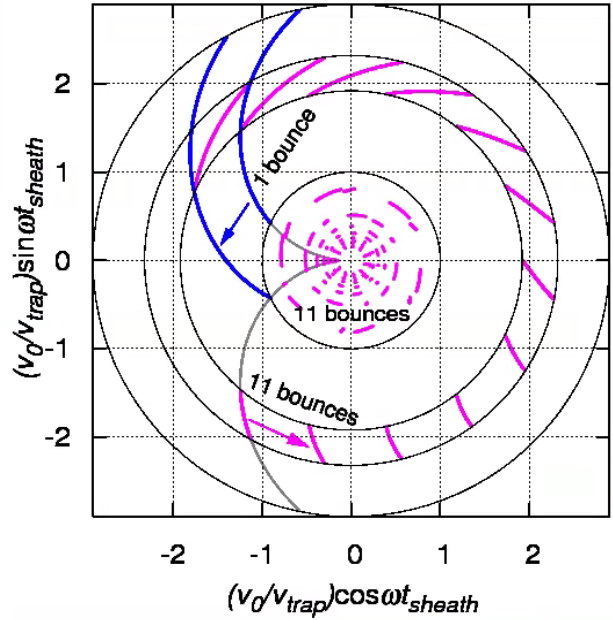

**Fig. 11: Orbits with 1 and 11 bounces off the DC sheath are selected. The 1-bounce orbit corresponds to the "main peak" in the detected EEPF. There is one 11-bounce orbit in in the "fast" region, and multiple such orbits in the "trapped" region.**

# IV. The role of electron-atom scattering collisions on EEDF at the RF electrode

In applications, RF-DC devices operate with electrode spacing of few centimeters and gas pressure in the range of 10–100 mTorr. The DC bias is on the order of several hundred volts, and $V_{pp}$ is in the kV range. To evaluate the influence of collisions on the EVDF of secondary electrons accelerated by the electrode sheaths, we compare the free path for electron-atom scattering with the electrode spacing $L$. For $L = 3$ cm and $p = 30$ mTorr such comparison is presented in Fig. 12. There, two horizontal lines at L=3 cm and L=6 cm show the lengths of single-pass (one-way) and double-pass (full-bounce) trajectories. The double pass distance (2L) becomes equal to the momentum relaxation length at $E \approx 300$ eV for $p = 30$ mTorr. Correspondingly, features that are associated with a single and double-pass should remain intact at these pressures. It should be noted that the energy determining the free path in the device is the energy gained in the sheath at moment of emission, and not the exit energy measured by the detector. Both the single-pass group (forming the ballistic peak at $E \approx V_{DC}$ ) and the double-pass group (emitted from the RF electrode and forming the next peak) have bouncing energies extending up to $V_{DC}$. Additional peaks, associated with multiple bounces inside gap due to trapping in between sheaths, will disappear as evident from Fig. 13 where the peak for single-bouncing electrons (emitted from the RF electrode and coming back), and the closely positioned step-like structure in the EVDF of originating from the DC cathode are still observed when collisions are taken into account. Although angular scattering for single- and double-pass populations is small, energy loss in inelastic collisions (predominantly ionization) needs to be taken into account in case of argon at $p \approx 50$ mTorr, when $\lambda \approx 5$ cm. The inelastic loss transfers electrons out of the peak into the tail below it on the energy scale.

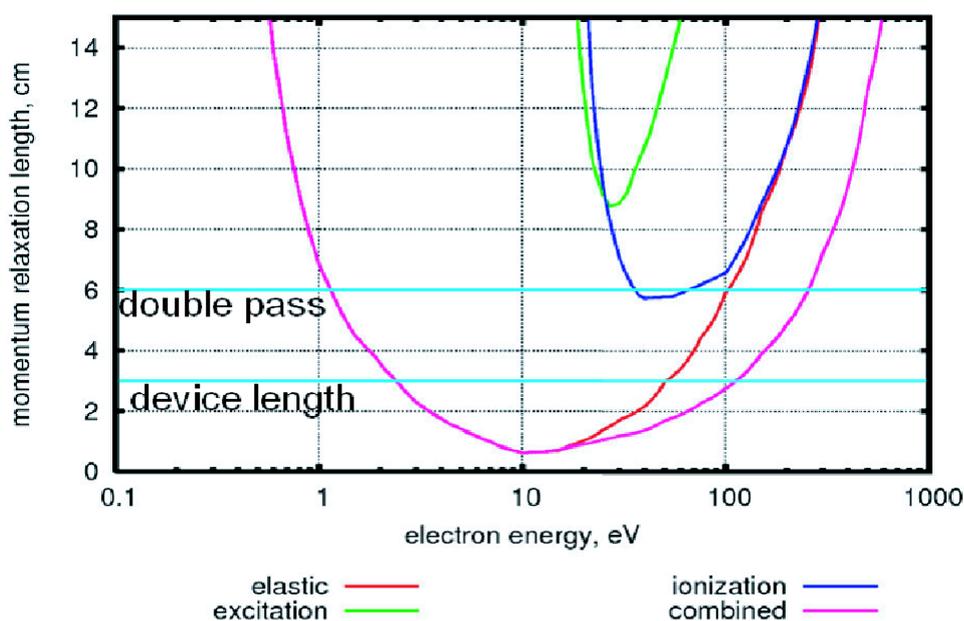

**Fig. 12: Momentum relaxation length vs. energy in argon at 30 mTorr.**

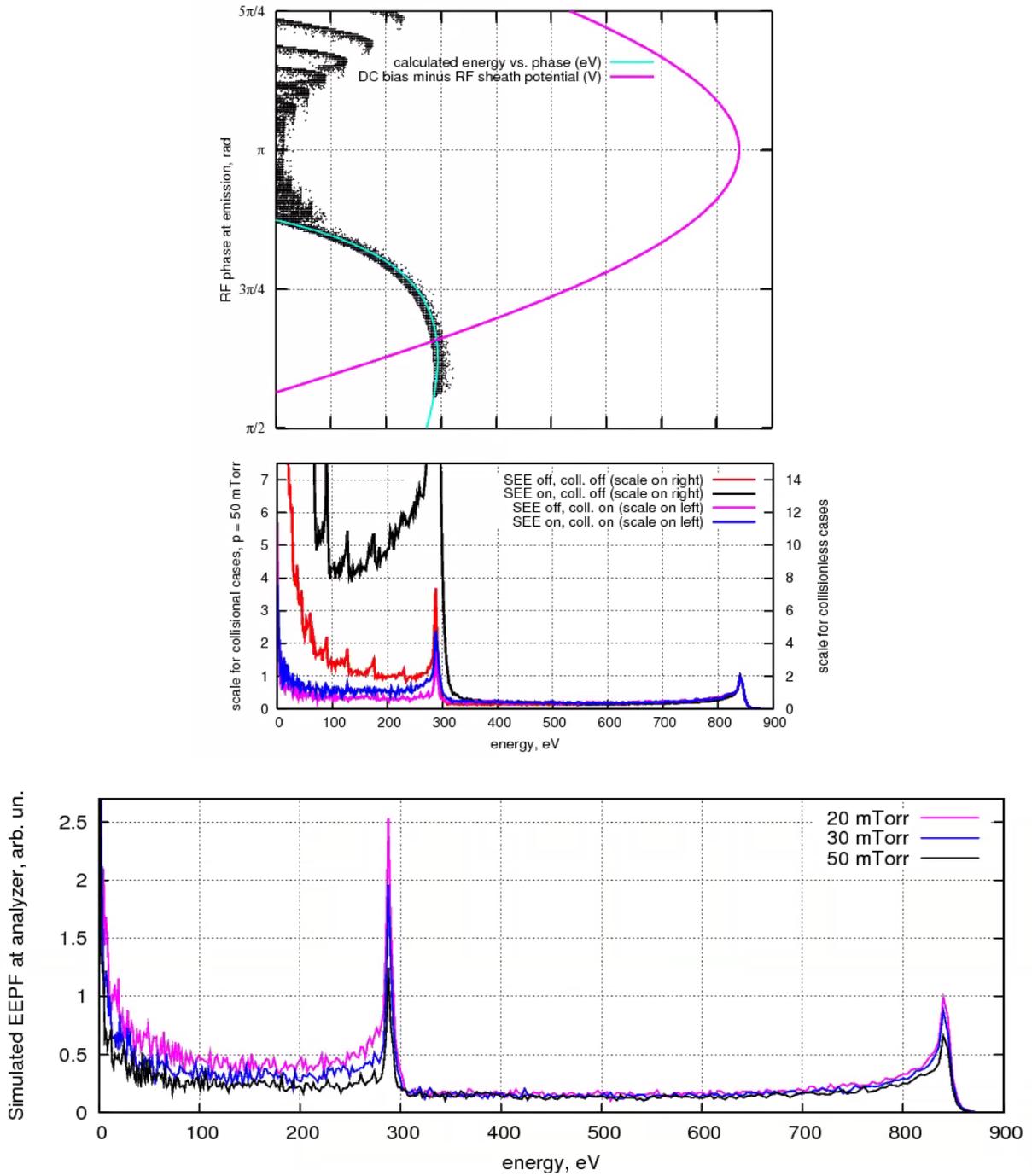

Fig. 13: Cycle-averaged Energy Probability Functions for electrons flowing out to the powered electrode, as would be seen by an energy analyzer, for $L$=3 cm, $V_{DC} = 800$ V and $V_{pp} = 2100$ V, in argon. Top : Effects of collisions (at 30 mTorr) and e-e emission (max. yield 1.2 at 400 eV). Bottom: Influence of gas pressure (e-e emission off) assuming constant ion-induced emitted fluxes.

The step-like structure is significantly weakened by collisions. Only those electrons can be collected by the detector that remain trapped for about one collision time, which corresponds only to about one double pass (single bounce) between the electrodes. Correspondingly, the EVDF jump at the step would be approximately two-fold instead of 5-fold as discussed in Sect. III.1. This is indeed seen, for $p$ = 30 mTorr, in test-particle simulations with account for collisions. In Fig. , simulated cycle-averaged EEPFs are compared with the results from the collisionless case. As expected, a high proportion of secondary electrons becomes trapped (and eventually thermalized) as a result of scattering, thus reducing the high-energy electron flux to the RF electrode several times compared to the idealized collision-free model.

The comparison also includes another simulation, which incorporated electron-induced secondary emission in addition to the constant injected flux representing the ion-induced emission current. Electron-electron emission was introduced based on Vaughan's parametrization of the yield curve, with $\gamma_{max} = 1.2$ at $E = 400$ eV, qualitatively suitable to represent the materials of both electrodes [8]. The influence of secondary e-e emission will be discussed in more detail in future work. Even at moderate level, it can strongly increase the density and current of the energetic electrons, due to finite lifetime of the confined population and to the multipactor effect.

In summary, even with account for collisions, the time-averaged EVDF or EEPF (the latter measured experimentally) of the electrons detected at the RF electrode should include peak and step-like features that enhance the distribution in the range of 200–400 eV. The presence of those features is due to the secondary electrons, emitted from both biased and RF-powered electrodes, which bounce at least once before exiting to the RF electrode.

## V. Effect of higher harmonics on EVDF at the RF electrode

The energy of secondary electrons emitted from the RF electrode and returning to it is sensitive to small higher-frequency content in the driving voltage waveform.   It will be shown that even single-bounce electron population can give rise to multiple peaks, which could happen in the collisional regime under the parameters realized in practice.   The basic justification for considering higher harmonics is, of course, that that the asymmetric discharge is a non-linear system and  $V(\phi)$  is close to purely sinusoidal only when the RF power supply is in the unloaded state.

It can be expected that a small high-frequency ripple in RF voltage would give rise to sufficiently strong oscillations in  $E(\psi)$  given by Eq.(27), resulting in extra maxima and minima and thus additional peaks in EVDF.   An example of this effect is shown in Fig. 9.

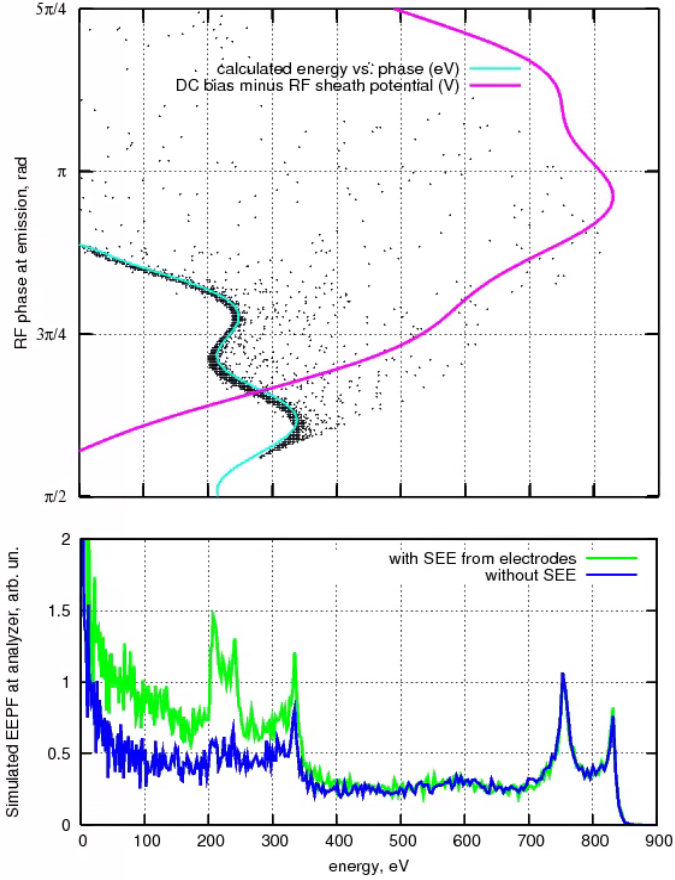

**Fig. 9: EVDF is sensitive to high harmonics. A 9th harmonic with amplitude of 3% of the fundamental has been added to the sinusoidal RF voltage. $V_{pp} = 2100$ V and $p = 50$ mTorr. The bottom panel now shows cycle-averaged EEPF collected within 25° cone in the velocity space, in analogy to the experiment.**

A 9th harmonic of the fundamental with amplitude of 3% of the fundamental cosine has been added to the sinusoidal RF voltage, to produce a waveform $\propto \{(1 + \cos\psi) - 0.03 \sin 9\psi\}$. This proves sufficient to create 3 energy peaks in the EEDF of the single-bouncing group, instead of one, because $E(\psi)$ exhibits strong oscillations, as evident in Fig. 10. The basic parameters were based on the experimental values of Ref. [6], namely $p = 50$ mTorr and $V_{pp} = 2100$ V. The EEPFs were simulated both with and without electron-induced secondary emission (SEE) from the powered and biased electrodes. The scatter plot in the top panel is for the case with SEE, and the dots situated away from the blue curve are due to elastic and inelastic reflections, accounted for in Vaughan's emission model (the envelope curve for the dots is $V_{DC} - V_{RF}[\psi + \tau(\psi)]$). We note that qualitatively, the resulting EEPF resembles the experimental result of Ref. [6] (their Fig. 2) and earlier results of Ref. [20]. This simple demonstration indicates that the shape of the RF waveform can be used to control the fine structure of the EEDF in the electron flux exiting the discharge. Finally, in Fig. 13, we show a phase-energy diagram for the high-harmonic case, idealized by removing the collisions. It should be compared with fig. 9.

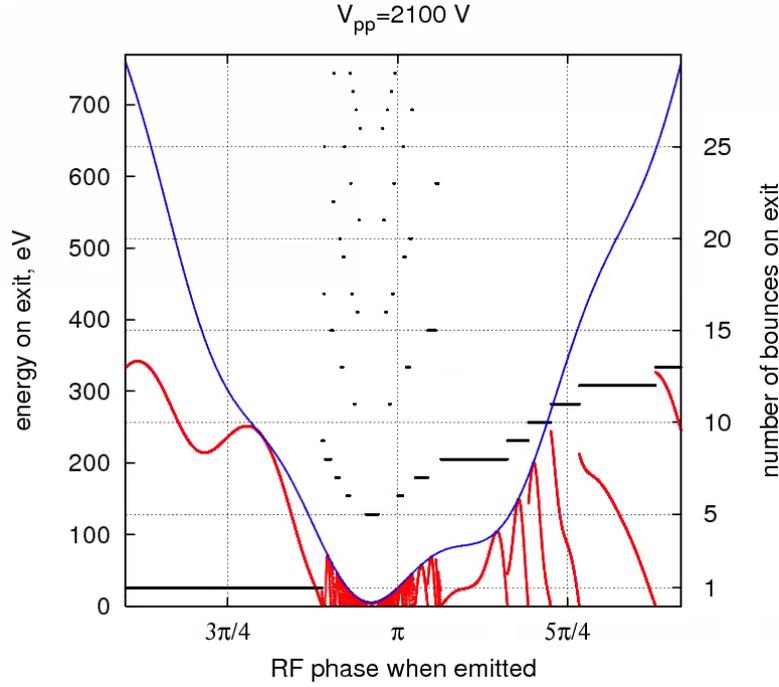

Fig. 10: Calculated phase-energy relationship for RF waveform with 9-th harmonic at 3% amplitude. Collisions have been switched off, but the minimum of RF sheath potential was 30 V, same as in test-particle runs. The blue curve is $V_{RF} - \min(V_{RF})$. It is seen that the bouncing energy for electrons forming each of the three peaks in the one-bounce group is above 200 eV and thus the peaks will not be completely erased by collisions at 50 mTorr or less.

## VI. Summary

The nature of electron velocity distributions in the bulk plasma and for the flux interacting with a wafer surface opposite a DC electrode appears at first glance of measured data difficult to explain. Apart from the "ballistic" peak (due to electrons travelling from the DC electrode without collisions and impinging on the opposing substrate) and the thermal population, a plethora of peaks are almost chaotic in nature. A number of possible explanations are possible for the nature and origin of all the peaks but a free-flight trap-and-release model of the accelerated secondary electrons in a hybrid RF-DC discharge is a logical starting point. In essence, the approach leverages the klystron effect in vacuum electronics. The main result is that the free-flight model predicts existence of peaks and step-like discontinuities in the energy distribution of the electrons impacting the RF electrode. By careful accounting of the kinematics of electrons through each pass and grouping them into trapped (by the potential well in the gap) and un-trapped, a straightforward theory can be used to identify the positions of peaks and steps, and the magnitudes of the steps. Such predictions are in good agreement with test particle simulations. The analysis proceeds with a collisionless approximation but we consider the effect of collisions between electrons and neutral atoms. Other effects which have been addressed, but remain to be fully explored, are secondary electrons generated at the DC and RF electrodes due to electron-electron emission (vs. ion-induced) and the presence of harmonics in the RF waveform. It is generally thought that collisions in the gap dampen fine structure in the EVDF. We show as is seen in experiment that certain structures survive in relevant pressure regimes, some of them due to secondary electrons produced by electron impact. Harmonics also provide an ample source to enhance middle-energy structure in the EVDF.

We note that the comparison with experiment is not exact and limited by the many approximations in our model and simulation. Comparisons are encouraging but we acknowledge the role of wave-particle interactions and bulk electron interactions with the sheath in playing potential roles in the formation of prominent EVDF structures observed in experiment. These will be the subject of companion publications.

This noted, this work points out that a simple analysis of electron motion that is occurring in the discharge can reproduce the essential features measured in experiment. We also observe that the experimental results should be very sensitive to the quality of the RF waveform and to the electron-impact secondary emission yield of the electrode materials.